\documentclass{vgtc}                         

\ifpdf%                                % if we use pdflatex
  \pdfoutput=1\relax                   % create PDFs from pdfLaTeX
  \usepackage{graphicx}                % allow us to embed graphics files
  \DeclareGraphicsExtensions{.pdf,.png,.jpg,.jpeg} % for pdflatex we expect .pdf, .png, or .jpg files
\else%                                 % else we use pure latex
  \usepackage{graphicx}                % allow us to embed graphics files
  \DeclareGraphicsExtensions{.eps}     % for pure latex we expect eps files
\fi%

\graphicspath{{figures/}{pictures/}{images/}{./}} % where to search for the images

\usepackage{microtype}                 % use micro-typography (slightly more compact, better to read)
\PassOptionsToPackage{warn}{textcomp}  % to address font issues with \textrightarrow
\usepackage{textcomp}                  % use better special symbols
\usepackage{balance}
\usepackage{mathptmx}                  % use matching math font
\usepackage{times}                     % we use Times as the main font
\usepackage{cite}                      % needed to automatically sort the references
\usepackage{tabu}                      % only used for the table example
\usepackage{booktabs}                  % only used for the table example
\newcommand\blfootnote[1]{%
  \begingroup
  \renewcommand\thefootnote{}\footnote{#1}%
  \addtocounter{footnote}{-1}%
  \endgroup
}

\onlineid{0}

\vgtccategory{Research}

\vgtcinsertpkg

\title{Designing interactive data visualizations representing recovery progress for patients after stroke}

\author{Alicia Ouskine \thanks{email: aliciaouskine@cmail.carleton.ca}\\ %
        \parbox{1.4in}{\scriptsize \centering Department of Human-Computer Interaction \\ Carleton University} %
\and Adrian D. C. Chan \thanks{email: adrian.chan@carleton.ca}\\ %
     \parbox{1.4in}{\scriptsize \centering Department of Systems and Computer Engineering \\ Carleton University \\ Affiliate Investigator at Bruyère Research Institute}
     \and Fateme Rajabiyazdi \thanks{email: fateme.rajabiyazdi@carleton.ca }\\ %
     \parbox{1.4in}{\scriptsize \centering Department of Systems and Computer Engineering \\ Carleton University}}

\abstract{ Stroke is one of the leading causes of disability worldwide. The efficacy of recovery is determined by a variety of factors, including patient adherence to rehabilitation programs. One way to increase patient adherence to their rehabilitation program is to show patients their progress that is visualized in a simple and intuitive way. We begin to gather preliminary information on Functional Capacity, Motor Function, and Mood/cognition from occupational Therapists at the Bruyere Hospital to gain a better understanding of how stroke recovery data is collected within in-patient stroke rehabilitation centers. The future aim is to design, develop, and evaluate a data visualization tool representing progress made by patients recovering from stroke.  }

\CCScatlist{
  \CCScatTwelve{Human-centered design}{Interactive Data-Visu\-al\-iza\-tion}{Visu\-al\-iza\-tion techniques}{Stroke}
  {Rehabilitation}
}

\begin{document}

\firstsection{Introduction}

\maketitle
There are over 50, 000 strokes in Canada annually (approximately one stroke every 10 minutes). Stroke is predicted to affect an increasing number of individuals due to population growth and aging. Stroke is a significant source of stress for patients, their families, and cost to the healthcare system. 

\blfootnote{
Poster presented at Graphics Interface Conference 2022\\
16-19 May - Montreal, QC, Canada\\
Copyright held by authors.\\ }

Patients after a stroke experience are often cared for by an interdisciplinary team that shapes a recovery program to improve motor function, postural control, and mobility. Adherence to these programs is key to recovery~\cite{duncan2002adherence}. However, an extended stay in a rehabilitation center can be tiring and frustrating for patients, and lack of motivation for goal-directed activities can reduce engagement and benefits from rehabilitation, impending stroke recovery. Tracking and reviewing recovery progress can give tangible feedback to keep patients motivated and reinforces adherence to programs.

Understanding complex recovery progress data collected over weeks or months can be demanding and difficult for patients and healthcare providers. An effective approach to overcoming this barrier is to visualize this recovery progress data. Visualizing health data can accurately show a summary of the data in an intuitive, simple, and accessible way~\cite{gotz2016data} to improve patients’ comprehension of their health status, increase engagement in care, and encourage adoption of positive health behaviors ~\cite{turchioe2019systematic}. 

In this study we gathered preliminary information through informal e-mail communication with health care providers that will help us design a interactive data visualization tool to increase engagement in patients recovering from stroke.

\section{Related Work}

%1. Visualization of health data (emr)
Visualizing medical data is a technique for organizing large amounts of data and extracting valuable information~\cite{vo2018electronic}. Data visualization for electronic medical records are frequently used in the development of clinical decision-support tools~\cite{rabbi2020reusable} and has garnered widespread interest due to its usefulness and significance~\cite{schneiderman2013}.  An example of an electronic medical record visualization is Lifelines~\cite{plaisant2003lifelines}. Lifelines is an interface that provides a visual overview and facilitates the navigation and analysis of clinical patient records. Visualizing medical data is not only valuable for healthcare providers but also to engage patients in their health management. For example, a study that explored patients’ and healthcare providers’ perspectives regarding presenting and reviewing patient data, discussed the importance of designing patient-generated visualizations for individuals to improve quality of life and to aid healthcare providers make decisions about patient ongoing care~\cite{rajabiyazdi:hal-02861239}.

%section 2. Rehab recovery data visualization
 
Data visualization that represents patient recovery allows for an increased understanding of patients health status, engagement, and adoption of positive health outcomes~\cite{turchioe2019systematic}. For example, AnatOnMe~\cite{Tao2011AnatOnMe}, is a projection-based handheld device designed to facilitate medical information exchange between healthcare providers and patients. AnatOnMe was found to increase patient engagement in their rehabilitation and understanding of medical information. The benefits from visualizing patient recovery data could be applied to increase patient engagement in stroke recovery. In a study that focuses on the use of an interactive dashboard prototype for upper limb movement information from stroke patients~\cite{ploderer2016therapists}, results show that therapists use the visualization of upper limb information to engage patients in the rehabilitation process through education, motivation, and discussion of experiences with activities of daily living, as well as to engage with other clinicians and researchers based on objective data~\cite{ploderer2016therapists}. The slow and gradual nature of stroke therapy, can make it difficult for stroke survivors to observe progress and can lead to frustration or lack of motivation towards goal-directed activities. Visualizing their rehabilitation progress could be a critical aspect for increasing patient involvement and motivation in their recovery.

%section 3. Stroke Recovery (this vis will be motivating.. --> gap - expanding elements of recovery data. )
Rehabilitation data visualizations has been shown to be beneficial to both patient and healthcare provider, however, there is a gap in literature on data visualizations representing patients overall recovery progress from stroke at in-patient rehabilitation facilities. We will fill this gap by taking preliminary steps to design and develop an interactive data visualization representing recovery after stroke.

% The visualization tool proposed in our study expand on elements of recovery data. The visualization tool will show rehabilitation progress of patients recovering from stroke in a simple and intuitive way to increase engagement in their rehabilitation program, and foster better patient-provider communication.

\section{Methods}
We started our study collecting information from a physiotherapist at a local long-term care Hospital to gain information on how stroke recovery data is obtained and stored. The physiotherapist then connected us with occupational Therapists and physiotherapists working in the unit. Through e-mail discussion, they were able to answer preliminary questions regarding data measuring patients recovering from stroke.

\section{Preliminary Results}

During our communication with physiotherapist and occupational therapist, we asked the following questions: 1. How often do you see patients after stroke? 2. What tools do you use to assess recovery for Functional Capacity, Motor Function, Mood/cognition? 3. How often do you use these assessment tools? 4. How do you store this information? 

Answers from occupational and physical therapists provided us with a starting point for understanding what assessment tools various healthcare providers use during a patient's stroke recovery process. Additionally, we formed a better understanding on how often the health care providers interact with patients recovering from stroke.

\section{Next Steps}
Our next step is to finalize ethics approval and coordination with the Bruyere rehabilitation center. Upon approval, we will retrieve and analyze anonymous retrospective data (e.g., vital signs, functional abilities, adherence to daily activities) collected from the patient recovering from stroke. Reviewing patient medical records is considered to be the gold standard to identify clinical data variables important to recovery. Additionally, we will conduct interviews with healthcare providers that have at least one year of experience working with patients recovering from stroke in order to gain an in-depth and rich understanding of the healthcare providers’ perspectives on how data measuring recovery is collected and used, and how to best represent patient recovery progress. Upon analyzing the results of our preliminary studies, in an iterative process, we will design and develop interactive data visualization tools displaying the patient recovery progress. Lastly, we will evaluate the visualization designs using a mixed-method approach by measuring patients’ adherence to rehabilitation programs through quantitative data such as participation in daily activities and conducting qualitative studies such as interviews with patients to gain their opinions of these data visualization tools and their effect on motivation.

\section{Conclusion}

We have currently retrieved preliminary information on assessment tools used by occupational and physical therapist to assess stroke recovery. We will further deepen this knowledge by interviewing physicians, nurses, occupational therapists, physical therapists, and speech pathologists that have more than one year experience working with patients recovering from stroke. Additionally, we will retrieve sample data from patients recovering from stroke. When enough information has been accumulated, we will design, develop, and evaluate the data visualization tool representing stroke recovery.

%% if specified like this the section will be committed in review mode
\acknowledgments{
This work was supported in part by the Natural Sciences and Engineering Research Council of Canada (NSERC) through the Collaborative Reseach and Training Experience (CREATE) and NSERC Discovery grants.}

\balance

\bibliographystyle{abbrv-doi}
\bibliography{references.bib}
\end{document}